\documentstyle[colordvi,psfig]{mn}   
\def\om{\Omega_p}
\def\omtw{\Omega_{\rm TW}}
\def\len{a_B}
\def\lag{D_L}
\def\vpd{{\cal R}}
\def\vpdtw{{\cal R}_{\rm TW}}
\def\kin{{\cal V}}
\def\pin{{\cal X}}
\def\sigkin{{\sigma_\kin}}
\def\sigpin{{\sigma_\pin}}
\def\wtslit{W_{\rm slit}}
\def\stdwgts{(\sigkin/\sqrt{N_{\rm slit}})^{-2}}
\def\padisc{PA$_{\rm disc}$}
\def\pabar{$\psi_{\rm bar}$}
\def\paell{$\psi_{\rm disc}$}
\def\paerr{$\delta_{\rm PA}$}
\def\epint{$\epsilon_{\rm D}$}
\def\omerr{\Delta\Omega/\om}
\def\scatter{\Delta_{\vpd}}

\def\scatterd{\Delta_{\vpd,\delta}}
\def\scattere{\Delta_{\vpd,\epsilon}}
\def\scatters{\Delta_{\vpd,spr}}
\def\gtsim{\mathrel{\spose{\lower.5ex \hbox{$\mathchar"218$}}
     \raise.4ex\hbox{$\mathchar"13E$}}}
\def\ltsim{\mathrel{\spose{\lower.5ex\hbox{$\mathchar"218$}}
     \raise.4ex\hbox{$\mathchar"13C$}}}

\def\degrees{^\circ}

\def\etal{{et al.}}
\def\eg{{\it e.g.}}
\def\etc{{\it etc.}}
\def\ie{{\it i.e.}}

\def\spose#1{\hbox to 0pt{#1\hss}}

\begin{document}   
   
\title[On Position Angle Errors in the Tremaine-Weinberg Method]
{On Position Angle Errors in the Tremaine-Weinberg Method}

   
\author[V.P. Debattista] 
{Victor P. Debattista\thanks{email: {\tt debattis@phys.ethz.ch}} \\
Astronomisches Institut, Universit\"at Basel, Venusstrasse 7,
CH-4102 Binningen, Switzerland\\ 
Current address: Institut f\"ur Astronomie, ETH H\"onggerberg, HPF G4.2,
CH-8093, Z\"urich, Switzerland
}

\date{{\it Draft version on \today}}

\maketitle   
   
\begin{abstract}   
  
  I show that Tremaine-Weinberg (TW) measurements of bar pattern speeds are
  sensitive to errors in the position angle of the disc, \padisc.  I use an
  $N$-body experiment to measure these errors; for typical random \padisc\ 
  errors, the resulting scatter in the measured values of the dimensionless
  bar speed parameter $\vpd$ (defined as the ratio of the corotation radius to
  the bar semi-major axis) is of order the observational.
  
  I also consider how the systematic \padisc\ errors produced by disc
  ellipticities affect TW measurements.  The scatter produced by these errors
  may be significant, depending on the ellipticity distribution.  Conversely,
  by using the sample of TW observations, I find that an upper limit of the
  typical disc (density) ellipticity is $0.07$ at the 90 per cent confidence
  level, which is in good agreement with previous measurements.
  
  Taken together, the random and systematic scatter suggest that the intrinsic
  distribution of $\vpd$ of gas-poor early-type barred galaxies may be as
  narrow as that of the gas-rich later-types.

\end{abstract}   
   

\section{Introduction}
\label{sec:intro}

Barred (SB) galaxies account for more than half of all high surface brightness
disc galaxies (Knapen 1999; Eskridge \etal\ 2000).  Recent observational and
theoretical studies have focused on the pattern speed of bars, $\om$.  The
quantity of greatest interest is $\vpd \equiv \lag/\len$, where $\lag$ is the
corotation radius and $\len$ is the semi-major axis of the bar.  A
self-consistent bar must have $\vpd \geq 1$ (Contopoulos 1980); bars with $1.0
\leq \vpd \leq 1.4$ are termed fast, while slow bars have larger $\vpd$.
Because bars have strong quadrupole moments, they lose angular momentum
efficiently in the presence of a dense dark matter halo (Weinberg 1985),
slowing down in the process; fast bars therefore have been interpreted as
evidence for maximum discs (Debattista \& Sellwood 1998, 2000, but see also
Valenzuela \& Klypin 2002).  Thus the accurate measurement of $\vpd$ in SB
galaxies is of interest.

Bar pattern speeds can be most reliably measured when kinematic data are
available.  One method relies on the dependence of the gas flow pattern on
$\om$, particularly at the shocks in the bar region.  Hydrodynamical
simulations can therefore recover $\om$; these find fast bars (\eg\ van Albada
\& Sanders 1982; Athanassoula 1992; Lindblad \& Kristen 1996; Lindblad \etal\ 
1996; Weiner \etal\ 2001).  An alternative method, which measures $\om$
directly, was developed by Tremaine \& Weinberg (1984).  Until now, the
Tremaine-Weinberg (hereafter TW) method has been applied to a small, but
growing, number of SB galaxies (published measurements are listed in Table
\ref{tab:tw_measurements}); all cases are consistent with fast bars.

Using 2-D absorption-line spectroscopy of the SB0 galaxy NGC 7079, Debattista
\& Williams (2003, in progress) show that the value of $\om$ obtained with the
TW method is sensitive to small errors in the position angle of the disc,
\padisc.  This raises the possibility that small errors in \padisc\ introduce
a significant scatter in TW measurements of $\vpd$.

Errors in \padisc\ can be either simple random ones, or systematic ones,
produced, for example, by deprojecting an intrinsically elliptical disc
assuming it is axisymmetric.
Constraints on the ellipticities\footnote{In this paper, disc ellipticity
  refers to the ellipticity, \epint, of the disc's density in its main plane.
  Expressions relating \epint\ and $\epsilon_\Phi$, the ellipticity of the
  potential in the disc plane, can be found in Franx \etal\ (1994).  Where the
  disc dominates the potential, \epint $>\epsilon_\Phi$} of discs come from a
variety of measurements.  The observed axes-ratios of galaxies show a deficit
of apparently circular discs, from which one concludes that perfect oblate
spheroids are poor fits to the data (Binney \& de Vaucouleurs 1981; Grosb\o l
1985).  Nevertheless, such studies find that typical ellipticities must be
small, \epint\ $\ltsim 0.1$ (Magrelli \etal\ 1992; Huizinga \& van Albada
1992; Lambas \etal\ 1992; Fasano \etal\ 1993).
Constraints on \epint\ are improved when kinematic data are included.  Rix \&
Zaritsky (1995) defined a sample of 18 kinematically face-on galaxies from the
Tully-Fisher relation (Tully \& Fisher 1977, hereafter the TF relation).
Using $K^\prime$-band photometry, they estimated typical $\epsilon_\Phi =
0.05^{+0.03}_{-0.02}$, with two arm spirals possibly accounting for some of
this signal.
Franx \& de Zeeuw (1992) showed that the small scatter in the TF relation
requires that $\epsilon_\Phi \leq 0.1$.  Since it is highly unlikely that all
the TF scatter is due to disc ellipticities alone, they concluded that a more
likely limit is $0 \leq \epsilon_\Phi \leq 0.06$.  By analysing the residuals
in the velocity-field of the gas ring around the S0 galaxy IC 2006, Franx
\etal\ (1994) found $\epsilon_\Phi = 0.012 \pm 0.026$ for this galaxy.
This approach has also been used by Schoenmakers \etal\ (1997) ($\epsilon_\Phi
< 0.1$ for 2 galaxies) and Beauvais \& Bothun (1999), (\epint\ $\ltsim 0.08$
for 6 galaxies).  An important uncertainty in this method is the viewing angle
of any ellipticity.  Andersen \etal\ (2001), therefore, measured \epint\ from
the discrepancies between photometric and kinematic disc parameters of nearly
face-on galaxies, finding an average \epint\ $=0.05$ for 7 galaxies; using the
same method on a larger sample of 28 galaxies, Andersen \& Bershady (2002)
were able to fit a log-normal distribution, with $\overline{\ln \epsilon_{\rm
    D}}\pm\sigma_{\ln\epsilon} = -2.82 \pm 0.73$ ($\overline{\epsilon_{\rm D}}
= 0.06^{+0.06}_{-0.03}$).  In all these studies, spirals may be responsible
for some or all of the signal seen (Barnes \& Sellwood 2003).
Finally, in the Milky Way Galaxy, a variety of constraints, local and global,
independently suggest $\epsilon_\Phi \simeq 0.1$, with the Sun close to the
minor-axis of the potential (Kuijken \& Tremaine 1994).

This paper studies the effect of \padisc\ errors on TW measurements.  In
Section \ref{sec:tw_method} I describe the TW method and its main sources of
uncertainty.  Most of these uncertainties can be quantified directly from
observations.  However, this is not generally true for errors due to \padisc\ 
uncertainties, so that some modelling is required.  Section \ref{sec:model}
therefore is devoted to setting up an $N$-body model for studying the impact
of \padisc\ errors on TW measurements.  In Section \ref{sec:randomerrs} I
demonstrate the sensitivity of the TW method to small \padisc\ errors and
estimate the scatter in $\vpd$ expected for the observational level of
\padisc\ uncertainty.  In Section \ref{sec:other_non_axi}, I consider the
scatter in $\vpd$ due to non-axisymmetric outer discs on TW measurements.  I
also obtain a novel constraint on \epint\ of early-type SB galaxies, based on
the requirement that none of the TW measurements thus far would have found a
value of $\vpd$ outside some range.  The result is in agreement with previous
determinations of \epint\ for unbarred galaxies.  In Section
\ref{sec:discussion}, I present my conclusions.  Throughout, I pay particular
attention to obtaining a conservative estimate of the scatter in $\vpd$ due to
\padisc\ errors.

\section{The TW method and its sources of errors}
\label{sec:tw_method}

The TW method requires a tracer population which satisfies the continuity
equation, and assumes that the time-dependence of the surface density,
$\Sigma$, can be expressed, in terms of cylindrical coordinates $(R,\phi)$ in
the disc plane, as:
\begin{equation}
\Sigma = \Sigma(R,\phi-\om t).  
\label{eqn:tw_condition}
\end{equation}
While not all non-axisymmetric structures obviously satisfy the condition of
equation \ref{eqn:tw_condition} (\eg\ warps), bars are well approximated by
this assumption.  The TW method is then contained in the following expression:
\begin{equation}
\pin \om = \kin/\sin i. 
\label{eqn:tw_eqn}
\end{equation}
Here, \mbox{$\pin = \int\,h(Y)\,X\,\Sigma\,dX\,dY$}, \mbox{$\kin =
  \int\,h(Y)\,V_{\rm los}\,\Sigma\,dX\,dY$}, $i$ is the disc inclination (I
use the convention $i = 0$ for face-on), $h(Y)$ is an arbitrary weighting
function, $V_{\rm los}$ is the line-of-sight velocity (minus the systemic
velocity) and $(X,Y)$ are galaxy-centered coordinates measured along the
disc's major (\ie\ inclination/line-of-nodes) and minor axes, respectively.
Equation \ref{eqn:tw_eqn} holds even when $\om = \om(t)$, as it must, since
the continuity equation is purely kinematic.

\begin{table}   
\caption{The sample of TW measurements in SB galaxies.  The references 
are: Kent 1987 (K87), Merrifield \& Kuijken 1995 (MK95), Gerssen \etal\ 1999
(GKM99)), Debattista \etal\ 2002a (DCA02) and Aguerri \etal\ 2003 (ADC03).  
The 6 galaxies from Debattista \etal\ (2002a) and Aguerri \etal\ (2003), which
have been analysed uniformly, constitute the ADC sample.}
\begin{center}   
\begin{tabular}{c c c c c}   
\hline   
\multicolumn{1}{c}{Galaxy} &   
\multicolumn{1}{c}{$i$} &     
\multicolumn{1}{c}{\pabar} &  
\multicolumn{1}{c}{$\vpd$} &     
\multicolumn{1}{c}{References} \\  
\hline   
 NGC 936 & $41\degrees$ & $66\degrees$ & $1.4 \pm 0.3$ & K87; MK95 \\
NGC 4596 & $38\degrees$ & $56\degrees$ & $1.2^{+0.4}_{-0.2} $ & GKM99 \\
NGC 1023 & $66\degrees$ & $78\degrees$ & $0.8^{+0.4}_{-0.2} $ & DCA02 \\
ESO 139-G009 & $46\degrees$ & $77\degrees$ & $0.8^{+0.3}_{-0.2} $ & ADC03  \\ 
IC 874       & $39\degrees$ & $70\degrees$ & $1.4^{+0.7}_{-0.4} $ & ADC03  \\ 
NGC 1308     & $36\degrees$ & $60\degrees$ & $0.8^{+0.4}_{-0.2} $ & ADC03  \\
NGC 1440     & $38\degrees$ & $40\degrees$ & $1.6^{+0.5}_{-0.3} $ & ADC03  \\ 
NGC 3412     & $55\degrees$ & $84\degrees$ & $1.5^{+0.6}_{-0.3} $ & ADC03  \\ 
\hline   
\label{tab:tw_measurements}   
\end{tabular}   
\end{center}   
\end{table}

Hydrodynamical studies find a narrow range in $\vpd = 1.2 \pm 0.2$.  The
quoted errors and spread in $\vpd$ when measured with the TW method are larger
(see Table \ref{tab:tw_measurements}).  Important sources of uncertainty in TW
measurements are:

\begin{enumerate}
  
\item{\it Uncertainty in $\om$.}  To obtain $\om$ with the TW method, the most
  commonly used strategy is to obtain several absorption-line slit spectra,
  for each of which $\kin$ and $\pin$ are measured.  Then plotting $\kin$
  versus $\pin$, one obtains $\om \sin i$ as the slope of the best-fitting
  straight line.  The values of $\pin$ are usually quite well defined; however
  values of $\kin$ tend to be noisy, and are the main source of uncertainty in
  $\om$.  This problem can be partly alleviated by projecting slit spectra
  along the spatial direction, thereby increasing the signal-to-noise ($S/N$)
  ratio (Merrifield \& Kuijken 1995).
  
\item{\it Uncertainty in $V_c$.}  Once $\om$ is measured, $\lag$ can be
  approximated as $V_c/\om$, where $V_c$ may be assumed flat.  However,
  because the tracer population must satisfy the continuity equation, the TW
  method is applied to early-type galaxies, which lack substantial patchy
  obscuring dust.  Unfortunately, their velocity dispersions are large, so
  that measurements of $V_c$ require correction for the asymmetric drift
  (unless gas is present outside the bar region [Gerssen 2002]).
  
\item{\it Uncertainty in $\len$.}  The bar semi-major axis is sometimes hard
  to measure in early-type galaxies since their bars often gradually blend
  into the disc.  The presence of massive bulges further complicates
  measurement of $\len$.
\end{enumerate}

For concreteness, note that the mean fractional uncertainties in $\om$, $V_c$
and $\len$ for the ADC sample (defined in Table \ref{tab:tw_measurements}) are
30, 7 and 19 per cent, respectively.  The resulting 67 per cent uncertainty in
$\vpd$, averaged over all the galaxies of Table \ref{tab:tw_measurements}, is
$\Delta_{\vpd, unc} = 0.7$.  (Meanwhile, the scatter of $\vpd$ for the full
sample, which includes both an observational error part and an intrinsic
distribution part, is $\Delta_{\vpd, obs} = 1.0$.  I measured this value by
using Monte-Carlo experiments in which I varied $V_c$ and $\len$ uniformly in
their error intervals, and varied $\om$ assuming its errors are Gaussian.)

Another source of error in the TW method is errors in the position angle of
the disc, \padisc.  Consider a slit observation: the right-hand side of
equation \ref{eqn:tw_eqn} then measures the flux of the tracer across the
slit.  However, this requires that the slit be exactly parallel to the $X$
axis; for any other orientation, the observed velocities do not measure the
full flux.  At the same time, $\pin$, the luminosity-weighted average position
along the slit, is rotated by the \padisc\ error.  The combination of these
two effects leads to an error in the measured $\om$.  Indeed, it is surprising
just how sensitive the TW method is to errors in \padisc: using 2-D
Fabry-Perot observations of NGC 7079, Debattista \& Williams (2003, in
preparation) show that errors of as little as $5\degrees$ in \padisc\ can lead
to errors in $\om$ of up to 100 per cent.  Published values of \padisc\ often
have uncertainties of this order.  While uncertainties in $\om$, $\len$ and
$V_c$ can be quantified directly from observations, errors in $\vpd$ due to
\padisc\ errors can only be modelled.

\section{Model and TW Measurements}
\label{sec:model}

\subsection{The $N$-body system}

\begin{figure}
\leavevmode{\psfig{figure=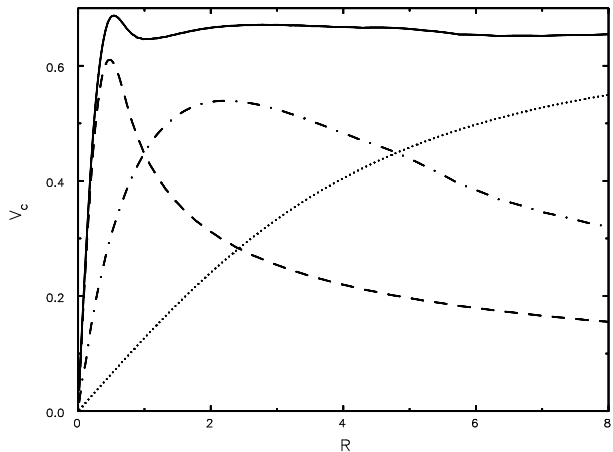,width=8.0truecm,angle=0}}
\caption[]{The initial rotation curve of the $N$-body model used.  
  The dashed, dot-dashed, and dotted lines represent the bulge, disc and
  frozen halo components respectively, while the solid line is the full
  rotation curve.}
\label{fig:rotcurv}
\end{figure}

\begin{figure}
\leavevmode{\psfig{figure=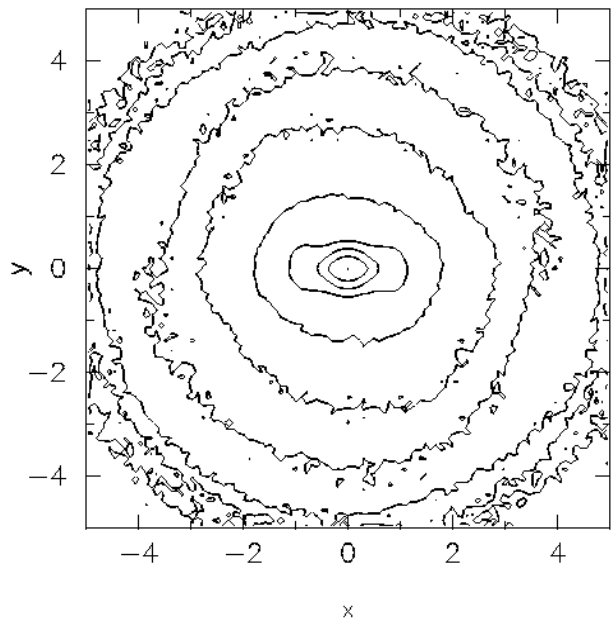,width=8.0truecm,angle=0}}
\caption[]{Logarithmically spaced contours of the disc $+$ bulge surface 
  density at $t=200$.  The system is rotating in the counter-clockwise sense.}
\label{fig:system}
\end{figure}

\begin{figure}
\leavevmode{\psfig{figure=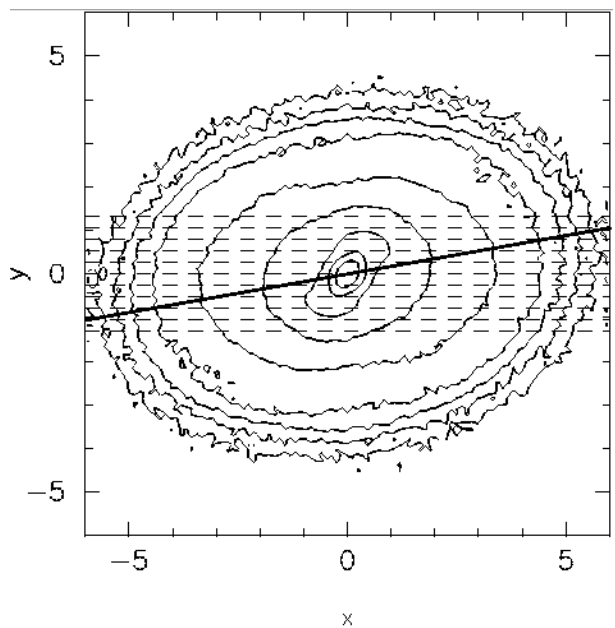,width=8.0truecm,angle=0}}
\caption[]{The system after rotation through \pabar\ $=60\degrees$, $i
  =45\degrees$ and \paerr $=+10\degrees$.  The solid line indicates the disc's
  true major axis, while the dashed lines indicate the (misaligned) slits
  used.}
\label{fig:example}
\end{figure}

\begin{figure}
\leavevmode{\psfig{figure=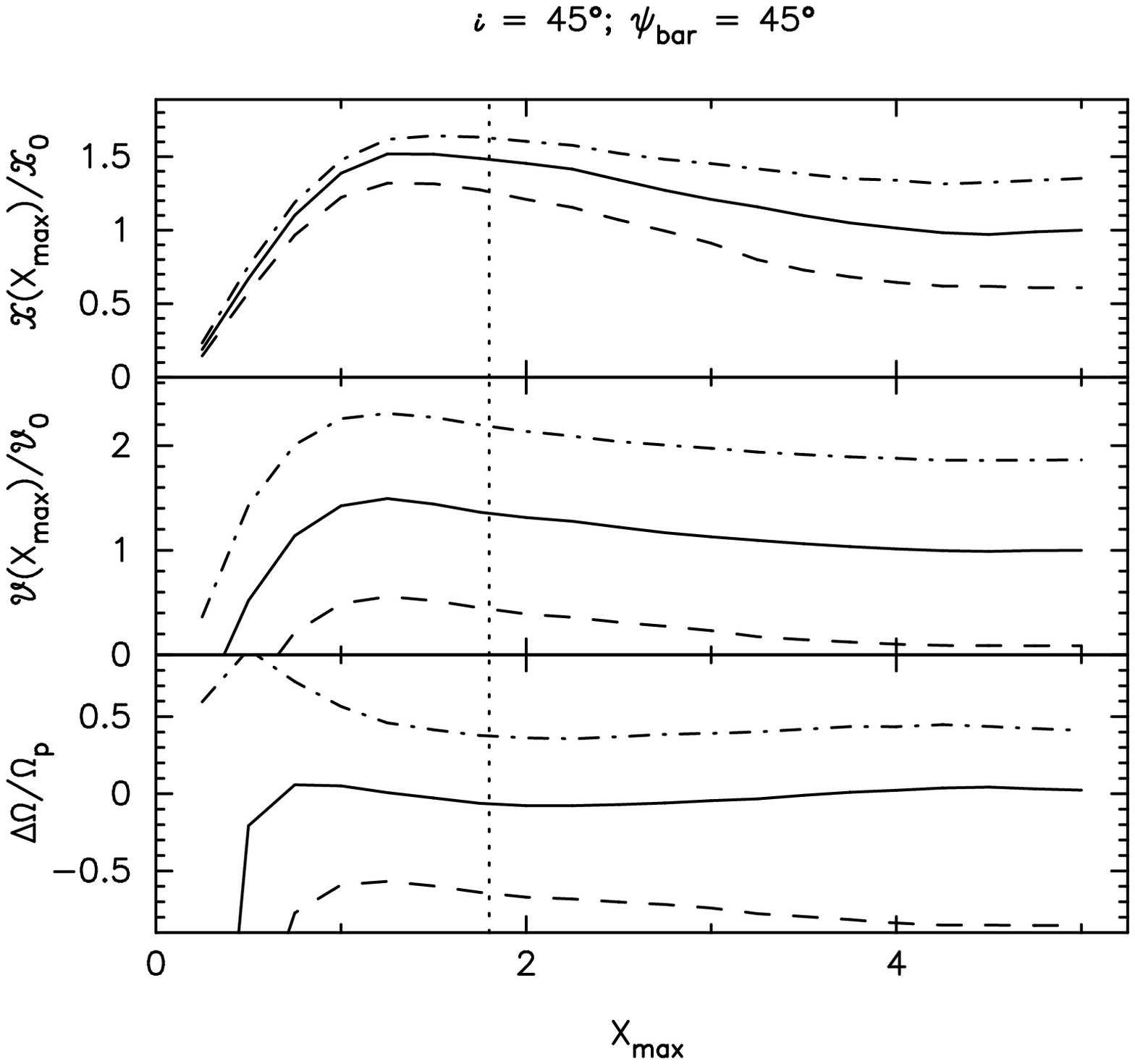,width=8.0truecm,angle=0}}
\caption[]{Variation of $\pin(X_{\rm max})$ (top) and $\kin(X_{\rm max})$ 
  (center) with $X_{\rm max}$.  Each line is normalized by the value of the
  full integral at \mbox{\paerr\ $=0$}.  The bottom panel shows the resulting
  fractional errors in the TW measurement of $\om$ using just this one slit.
  In all panels, the solid, dashed and dot-dashed lines are for \mbox{\paerr\ 
    $=0$}, \mbox{\paerr\ $=-5\degrees$} and \mbox{\paerr\ $=+5\degrees$},
  respectively.  The dotted vertical lines indicate $\len$.  Other values of
  \pabar\ and $i$ give qualitatively similar results.}
\label{fig:vandxwithxmax}
\end{figure}

\begin{figure}
\leavevmode{\psfig{figure=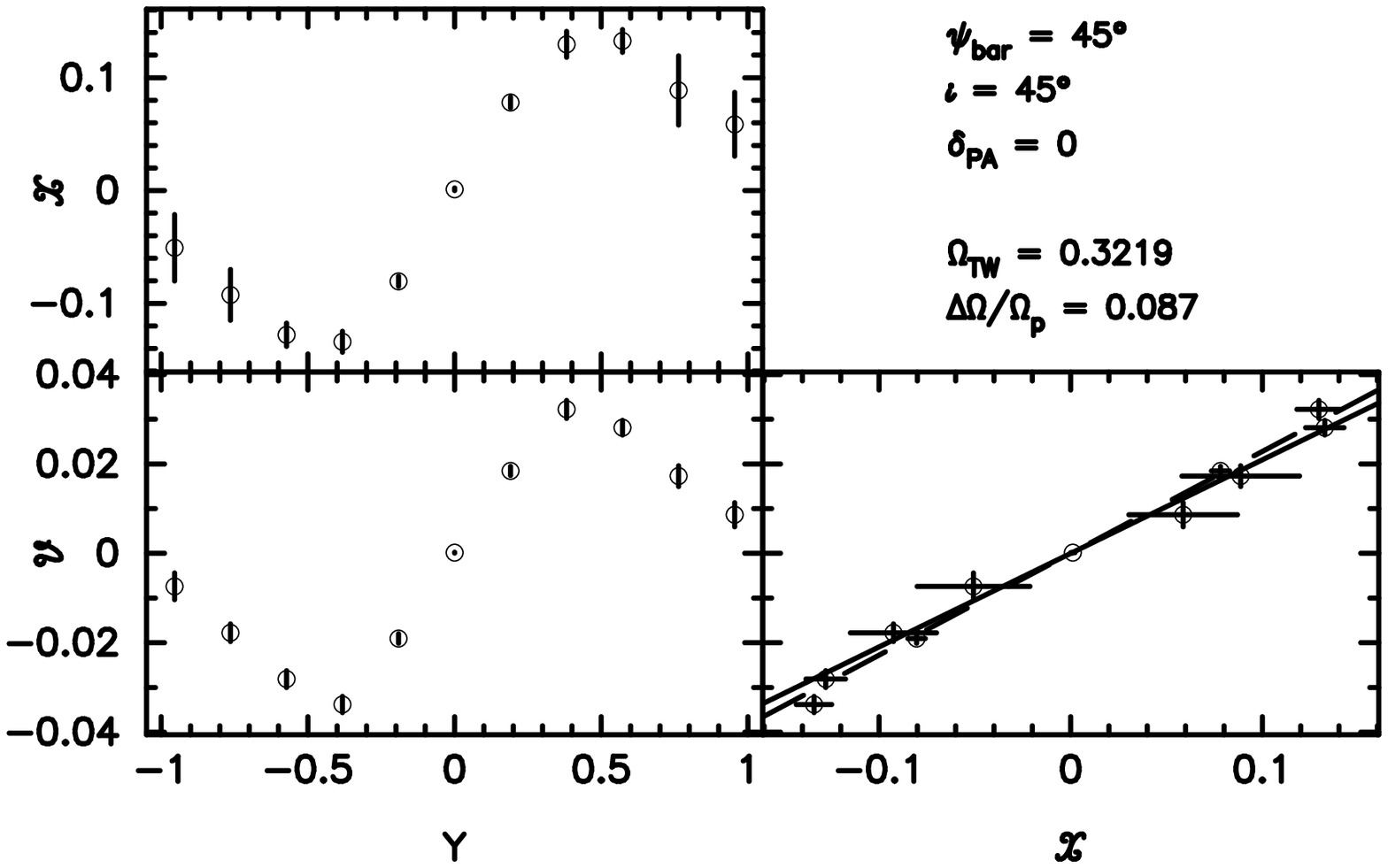,width=8.0truecm,angle=0}}
\caption[]{The TW integrals for \pabar $=45\degrees$, $i=45\degrees$.  On
  the left are shown $\pin$ (top) and $\kin$ (bottom) as functions of the slit
  offset.  On the right, $\kin$ is plotted against $\pin$, and a straight line
  fit.  The solid line has slope $\om\sin i$ as measured from the time
  evolution, while the dashed line shows the best-fitting straight line, with
  slope $\omtw \sin i$.  Each slit contains $\gtsim 10^5$ particles; all
  errors have been enlarged by a factor of 1000 for clarity.}
\label{fig:signals}
\end{figure}

\begin{figure}
\leavevmode{\psfig{figure=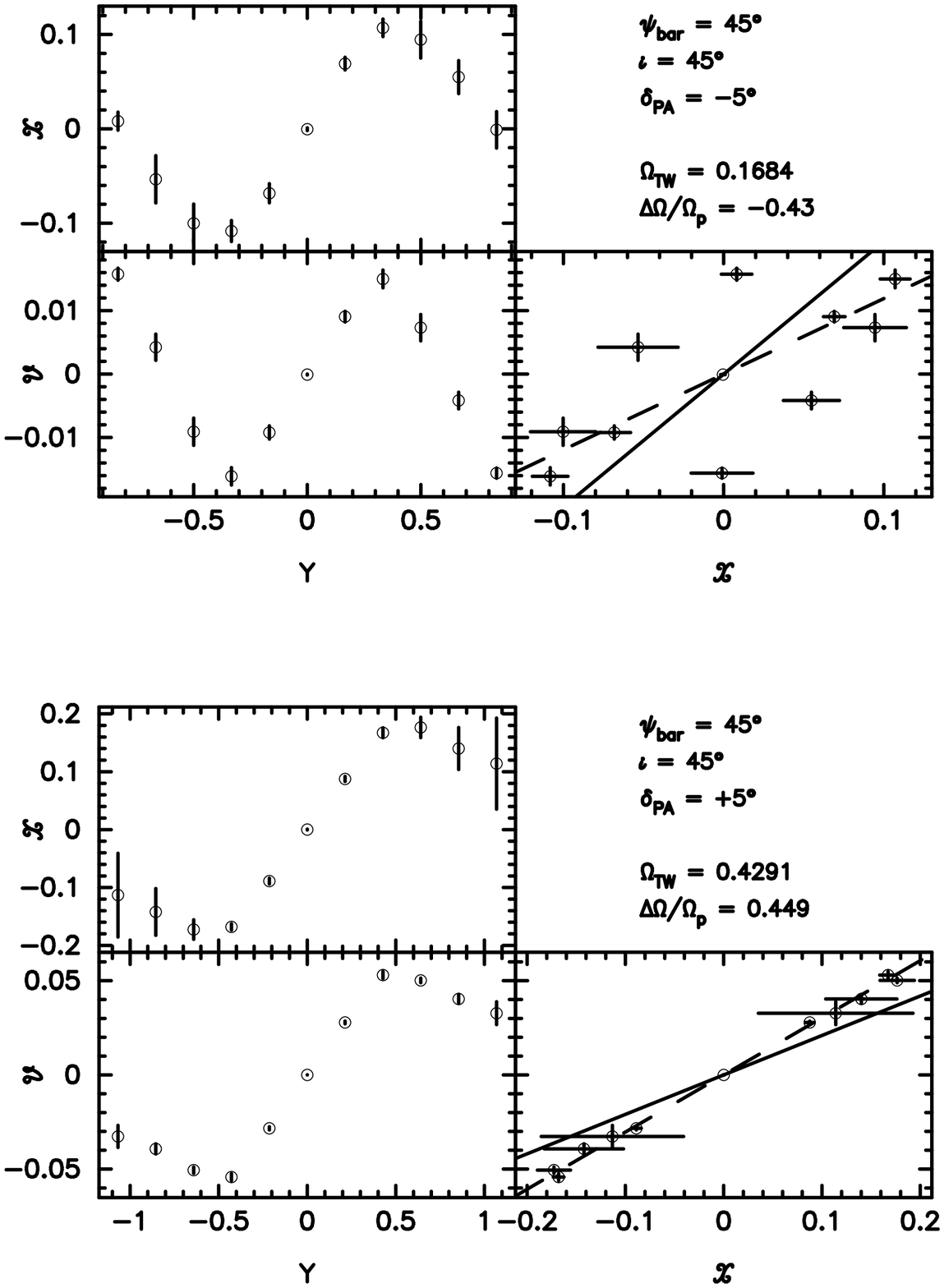,width=8.0truecm,angle=0}}
\caption[]{As in Fig. \ref{fig:signals} but for \paerr\ $=-5\degrees$
  (top 3 panels) and \paerr\ $=+5\degrees$ (bottom 3 panels).}
\label{fig:signals2}
\end{figure}

To quantify better the sensitivity of the TW method to errors in \padisc, I
applied it to a high resolution $N$-body bar.  In numerical simulations, $\om$
can be measured accurately directly from the time evolution, which makes
possible a comparison with TW measurements at various disc and bar
orientations and errors in \padisc.
The simulation which produced the model of an early-type galaxy consisted of
live disc and bulge components inside a frozen halo.  The frozen halo was
represented by a spherical logarithmic potential
\begin{equation}
\Phi_L(r) = \frac{1}{2} v_0^2 \ln(r_c^2 + r^2),
\end{equation}
where $r_c$ is the core-radius and $v_0$ is the asymptotic circular
velocity.  
The initially axisymmetric disc was modelled by an exponential disc with a
Gaussian thickening
\begin{equation}
\rho_{\rm d}(R,z) = \cases{ \displaystyle 
f_{\rm d} \frac{M}{2 \pi R_{\rm d}^2} e^{-R/R_{\rm d}}
\frac{1}{\sqrt{2 \pi} z_{\rm d}}e^{-\frac{1}{2}(z/z_{\rm d})^2} &
      $R \leq R_{\rm t},$ \cr
     \cr
  0 & $R > R_{\rm t}$, \cr}
\end{equation}
where $f_{\rm d}$ is the fraction of the active mass which is in the disc and
$R_t$ is the radius at which the disc is truncated.
The bulge was generated using the method of Prendergast \& Tomer (1970), where
a distribution function is integrated iteratively in the global potential,
until convergence.  For this application, I used the distribution function of
a lowered, $n=2$, polytrope, truncated at $r_{\rm b}$
\begin{equation}
f(\bmath{x},\bmath{v}) = C {\cal F}(E) = 
C \left\{ \left[-2 E(\bmath{x},\bmath{v}) \right]^{1/2} - 
          \left[-2E_{\rm max}\right]^{1/2} \right\}.
\label{eqn:df}
\end{equation}
Here $C$ is a mass normalization constant and $E_{max} = \Phi_{tot}(r_{\rm
  b})$, the total potential at $r_{\rm b}$ in the disc plane.  Disc kinematics
were set up using the epicyclic approximation to give Toomre $Q = 2.5$, a
value appropriate for an early-type disc galaxy; this leads to weak spirals,
which do not interfere substantially with measurements of $\om$.  Vertical
equilibrium was obtained by integrating the vertical Jeans equation.  The disc
and bulge were represented by $4\times 10^6$ equal-mass particles, giving a
mass ratio $M_{\rm d}:M_{\rm b} = f_{\rm d}:1-f_{\rm d} = 0.8:0.2$.  Further
details of the setup methods used can be found in Debattista \& Sellwood
(2000).

In units where $R_{\rm d} = M = G = 1$, which gives a unit of time $(R_{\rm
  d}^3/GM)^{1/2}$, the values chosen for the various parameters are given in
Table \ref{tab:run216}.  This choice of parameters gives a flat rotation curve
out to large radii, as shown in Fig.  \ref{fig:rotcurv}.

The simulation was run on a 3-D cylindrical polar grid code (described in
Sellwood \& Valluri [1997]) with $N_R\times N_\phi \times N_z = 60 \times 64
\times 225$.  The radial spacing of grid cells increases logarithmically from
the center, with the outer edge of the grid at just over $15 R_{\rm d}$.  The
vertical spacing of the grid planes, $\delta z$, was set to $0.0125 R_{\rm
  d}$.  I used Fourier terms up to $m=8$ in the potential, which was softened
with the standard Plummer kernel, of softening length $\epsilon = 0.0125
R_{\rm d}$.  Time integration was performed with a leapfrog integrator using a
fixed time-step $\delta t = 0.02$.

The equilibrium set up using epicyclic theory is rather approximate at this
high $Q$; nonetheless, the system quickly relaxes to a new equilibrium close
to the initial conditions.  The resulting axisymmetric system is unstable and
forms a rapidly rotating bar by $t=150$.  Fig. \ref{fig:system} shows the
system at $t=200$, the time I chose for this analysis; by this time, the bar
had gone through a period of growth and $\om$ had settled to a well defined
value.  The bar is strong in the disc, with a weaker triaxiality in the bulge.
The values of the bar's parameters at this time are given in Table
\ref{tab:run216}.  Note that the resulting $N$-body model of an SB0 galaxy is
reasonable, with a bar which is neither too weak nor too strong, having
$\len/R_{\rm d}$ towards the upper limit of, but within, the range of the ADC
sample.

Since the dark matter halo is frozen, $\om$ remains constant except for small
oscillations produced by interference with weak spirals.  I chose $t=200$
because the spirals were relatively weak at this time, allowing me to measure
$\om$ with a minimum of interference.

\begin{table}
\vbox{\hfil
\begin{tabular}{ rcl }
\hline
& $t=0$ & \\
\hline
                  Halo core radius &  & $r_c = 5$ \\
            Halo circular velocity &  & $v_0 = 0.648$ \\
                  Disc scaleheight &  & $z_{\rm d} = 0.1$ \\
            Disc truncation radius &  & $R_t = 5$ \\
           Bulge truncation radius &  & $r_{\rm b} = 0.78$ \\
\hline                               
& $t=200$ & \\    
\hline                               
               Bar semi-major axis &  & $\len = 1.8 \pm 0.1 $ \\
                 Bar pattern speed &  & $\om = 0.296 \pm 0.011 $ \\
               Bar speed parameter &  & $\vpd = 1.2 \pm 0.1 $ \\
\hline
\end{tabular}
\hfil}
\caption{Parameter values of the $N$-body model.}
\label{tab:run216}   
\end{table}

\subsection{Pattern speed measurements}

For TW measurements on the $N$-body system, I began with the disc in the
$xy$-plane with the bar along the $x$-axis, as in Fig. \ref{fig:system}.  For
an observer at positive $z$, viewing the system at an arbitrary orientation
requires three rotations.  Rotating the system (rather than the frame), the
first rotation is about the $z$-axis through an angle \pabar, followed by a
rotation about the $x$-axis to give an inclination $i$.  At this point, the
$XY$ frame of the TW integrals is identical to the $xy$ frame.  A third
rotation, through an angle \paerr\ about the $z$-axis, introduces an error in
\padisc\ if the observer continues to identify $(X,Y)$ with $(x,y)$.  (Note
that, in this definition, \paerr\ $>0$ moves the assumed disc major-axis away
from the bar's major-axis.)  From here on, for notational convenience, I refer
to the $X$ and $Y$ axes as the {\it assumed} major and minor axes of the
system (\ie\ the $x$ and $y$ axes), even when \paerr\ $\neq 0$.  Fig.
\ref{fig:example} shows an example of the system after such a series of
rotations.

I measured $\pin$ and $\kin$ for $0\degrees \leq$ \pabar\ $\leq 90\degrees$,
$0\degrees \leq i \leq 90\degrees$ and $-90\degrees \leq $ \paerr\ $\leq
90\degrees$ in 11 slits covering the region $-Y_{max} \leq Y \leq Y_{max}$.
Here $Y_{max}$ is $1.2 \times$ the largest of the projections onto the
$Y$-axis of the bar's 3 principal axes.  This limited range in $Y$ mimics the
typical observational setup, and reduces the noise in the measurement.  The
values of $\pin$ and $\kin$ for each slit were obtained as:
\begin{equation}
\pin = \frac{1}{{\cal P}}\sum_{i \in {\rm slit}}w_i X_i, \ \ \ 
\kin = \frac{1}{{\cal P}}\sum_{i \in {\rm slit}}w_i V_{z,i},
\end{equation}
where $V_{z,i}$ and $X_i$ are the line-of-sight velocity and $X$ coordinate of
particle $i$, $w_i$ is the weight assigned to each particle and ${\cal P} =
\sum_{i \in {\rm slit}}w_i$ (which corresponds to $h(Y) = 1/\int \Sigma dX$,
so that $\pin$ and $\kin$ are the luminosity-weighted average position and
velocity of each slit, as in observations).  Except where noted, I used $w_i =
1$ for all particles, whether disc or bulge; thus ${\cal P} = N_{\rm slit}$,
the number of particles in the slit.  If $\pin(X_{\rm max})$ and $\kin(X_{\rm
  max})$ represent the integrals extending from $-X_{\rm max}$ to $X_{\rm
  max}$, then error estimates $\sigpin$ and $\sigkin$ were obtained by
considering their maximum variation with $X_{\rm max}$ outside the bar radius.
Because the number of particles in each slit was high, these radial variations
are due only to weak non-axisymmetric structure at large radius.  In Fig.
\ref{fig:vandxwithxmax}, I show $\pin(X_{\rm max})$ and $\kin(X_{\rm max})$
for a typical slit.

To measure the pattern speed from a set of such slits, I fit a straight line
to $\kin$ as a function of $\pin$, as in observations, using least-squares
weights $\wtslit$.  The principal observational uncertainty is in $\kin$ and
is due to photon statistics; I therefore used $\wtslit = \stdwgts$.

The slope of this fitted line is $\omtw \sin i$, where I use the notation
$\omtw$ to distinguish from the pattern speed measured through the time
evolution.  An example of such a fit is shown in Fig. \ref{fig:signals}, which
reveals that $|\pin|$ and $|\kin|$ increase with increasing $|Y|$, until they
reach a maximum, and then decrease.  Observational requirements of high $S/N$
in modest time usually restricts slit offsets to ones at, or inside, the
maximum in $|\pin|$ (\eg\ Aguerri \etal\ 2003).

I verified that the TW method accurately measures $\om$ when \paerr\ $=0$: in
the range $10\degrees \leq i \leq 80\degrees$ and $10 \degrees \leq$ \pabar\ 
$\leq 80\degrees$, fractional errors, $|\omerr| \equiv |(\omtw - \om)/\om|$,
are smaller than 20 per cent, in agreement with Tremaine \& Weinberg (1984).

Besides this experiment, I tried various others.  For example, in two
experiments, I set $w_i = 0$ and $w_i = 2$ for the bulge particles, leaving
$w_i = 1$ for the disc ones.  The results were consistent with those presented
above, leading me to conclude that any plausible difference between the
stellar mass-to-light ratio of the bulge and disc does not introduce large
errors in $\omtw$.

\section{Simple \padisc\ errors}
\label{sec:randomerrs}

\subsection{Sensitivity to errors in \padisc}

Fig. \ref{fig:vandxwithxmax} also plots $\pin(X_{max})$ and $\kin(X_{max})$
for \paerr\ $=\pm5\degrees$.  It is clear that these small errors in \padisc\ 
change the values of $\pin(X_{max})$ and $\kin(X_{max})$ substantially, while
qualitatively looking similar to the \paerr\ $=0$ case.  Moreover, these
changes are at all $X_{max}$, particularly in the case of $\kin(X_{max})$;
thus, limiting the integrals to small $X_{\rm max}$ does not diminish the
error (although it does not increase it, either, unless $X_{max}$ is well
within the bar).  For this one slit, these changes gave an $\omtw$ which is in
error by up to 100 per cent.

\begin{figure}
\leavevmode{\psfig{figure=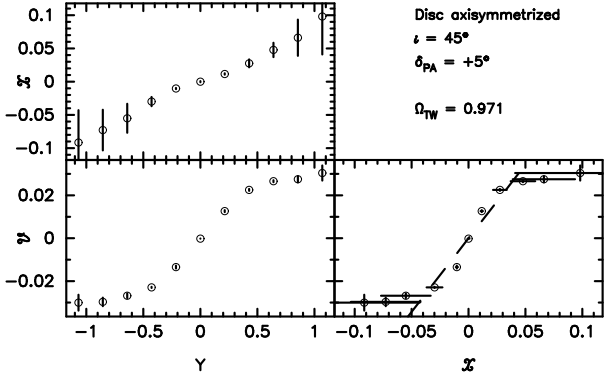,width=8.0truecm,angle=0}}
\caption[]{As in Fig. \ref{fig:signals} but for \paerr\ $=+5\degrees$
  with the disc axisymmetrized by shuffling the particles in azimuth.  The
  pattern speed fit in the right panel is, therefore, merely an artifact.}
\label{fig:axisym}
\end{figure}

In Fig. \ref{fig:signals2}, I again plot the integrals as a function of $Y$,
but this time for \paerr\ $=\pm5\degrees$.  Both $|\pin|$ and $|\kin|$ reach a
smaller (larger) maximum in the case of \paerr\ $=-5\degrees$ (\paerr\ 
$=+5\degrees$), while at larger offsets, the decrease in the values of the
integrals is faster (slower) than in the \paerr\ $=0$ case; for \paerr\ 
$=-5\degrees$, $\kin$ even switches sign.

To begin to understand these changes, I consider an axisymmetric system.  For
a slit at $Y>0$, when \paerr\ $=0$, the contribution to $\pin$ and $\kin$ from
$-X$ is exactly cancelled by that from $+X$.  When \paerr\ $>0$, several
changes occur.  First, $+X$ is always closer to the galaxy center (in the
disc's own plane), and at a smaller angle from the intrinsic major-axis, than
is $-X$.  Therefore $|V_{\rm los}(+X)| > |V_{\rm los}(-X)|$, if the rotation
curve is flat, giving $\kin$ a positive perturbation, which is further
enhanced if the density profile of the disc decreases radially, as is
generally the case.  The changes in $\pin$ are due solely to the radial
variation of the surface density; when this is constant everywhere, $\pin$ is
exactly zero at all \paerr.  Conversely, an exponential disc with small
scale-length (relative to the slit offset) gives large values of $\pin$ when
\paerr\ $\neq 0$.  The change in $\kin$ is large already at small $X_{\rm
  max}$ (see Fig.  \ref{fig:vandxwithxmax}), whereas the changes in $\pin$ are
more distributed over $X_{\rm max}$.  This behavior is due to the fact that
the integrand $\Sigma V_{\rm los}$ grows more rapidly with $X$ than does
$\Sigma X$.  Indeed, for a flat rotation curve $|V_{\rm los}(X) + V_{\rm
  los}(-X)|$ is largest at $X=0$.

Fig. \ref{fig:axisym} plots $\pin$ and $\kin$ at \paerr\ $=+5\degrees$ for the
axisymmetric disc produced by randomizing the azimuthal coordinate of all the
particles in the $N$-body model (preserving the average radial density
profile).  Even in the absence of any non-axisymmetric structure, misaligned
slits produce non-zero $\pin$ and $\kin$, which may plausibly be fit to a
pattern speed where none is present.

These extra contributions to $\pin$ and $\kin$ will still be present in the
barred case, modified by the presence of the bar (\eg\ $\pin$ will still
change even when the azimuthally averaged radial profile is constant, and
$\pin$ changes sign if the bar crosses the $Y$-axis), but fundamentally of the
same character.  It is then easy to imagine that some combination of Fig.
\ref{fig:signals} and Fig. \ref{fig:axisym} produces the bottom panels of Fig.
\ref{fig:signals2}, at least qualitatively.  For \paerr\ $=-5\degrees$, the
signs of $\pin$ and $\kin$ in Fig. \ref{fig:axisym} would be reversed, which
then combines with Fig. \ref{fig:signals} to produce something like the top
panels of Fig. \ref{fig:signals2}.

Fig. \ref{fig:signals2} suggests that, when \paerr\ $<0$, it may be possible
to recognize \paerr\ $\neq 0$ by the large $\chi^2$ in the linear regression.
Unfortunately the most discrepant points are the ones at large offset; in
observations, their $\sigkin$ will certainly be (fractionally) much larger
than here, in which case $\chi^2$ is not likely to be greatly increased by
these points.  Moreover, the two most discrepant points are at small $|\pin|$
and are thus unlikely to have been chosen for observation in the first place.
It therefore seems likely that, in the absence of considerable investment in
telescope time (which anyway would not catch \paerr\ $>0$), the error in
\padisc\ would go unnoticed.

\begin{figure}
\leavevmode{\psfig{figure=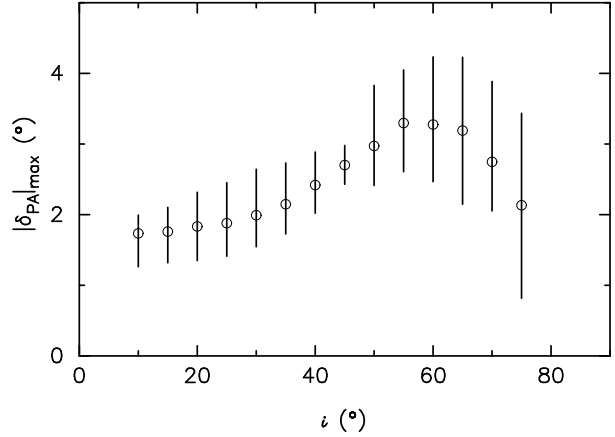,width=8.0truecm,angle=0}}
\caption[]{The maximum permitted error in \padisc\ required for $\omtw$ 
  accurate to 30 per cent.  These have been computed in the range $15\degrees
  \leq $ \pabar\ $ \leq 75\degrees$, with circles representing the averages
  and the error bars indicating the extreme cases.}
\label{fig:thirty}
\end{figure}

The $5\degrees$ errors of Fig. \ref{fig:signals2} give errors in $\omtw$ as
large as 48 per cent.  In Fig. \ref{fig:thirty}, I present the largest errors
permitted to guarantee $\omtw$ accurate to 30 per cent.  The limits on \paerr\ 
are quite stringent: $|$\paerr$| \leq 4\degrees$ is needed at $i=60\degrees$
and the limit is smaller at other inclinations.  (Note, however, that for
$|\omerr|$ to be larger than 30 per cent, it is necessary, {\it but not
  sufficient} for $|$\paerr$|$ to be larger than the values given in Fig.
\ref{fig:thirty}, since \paerr\ can be either positive or negative.)

\begin{figure}
\leavevmode{\psfig{figure=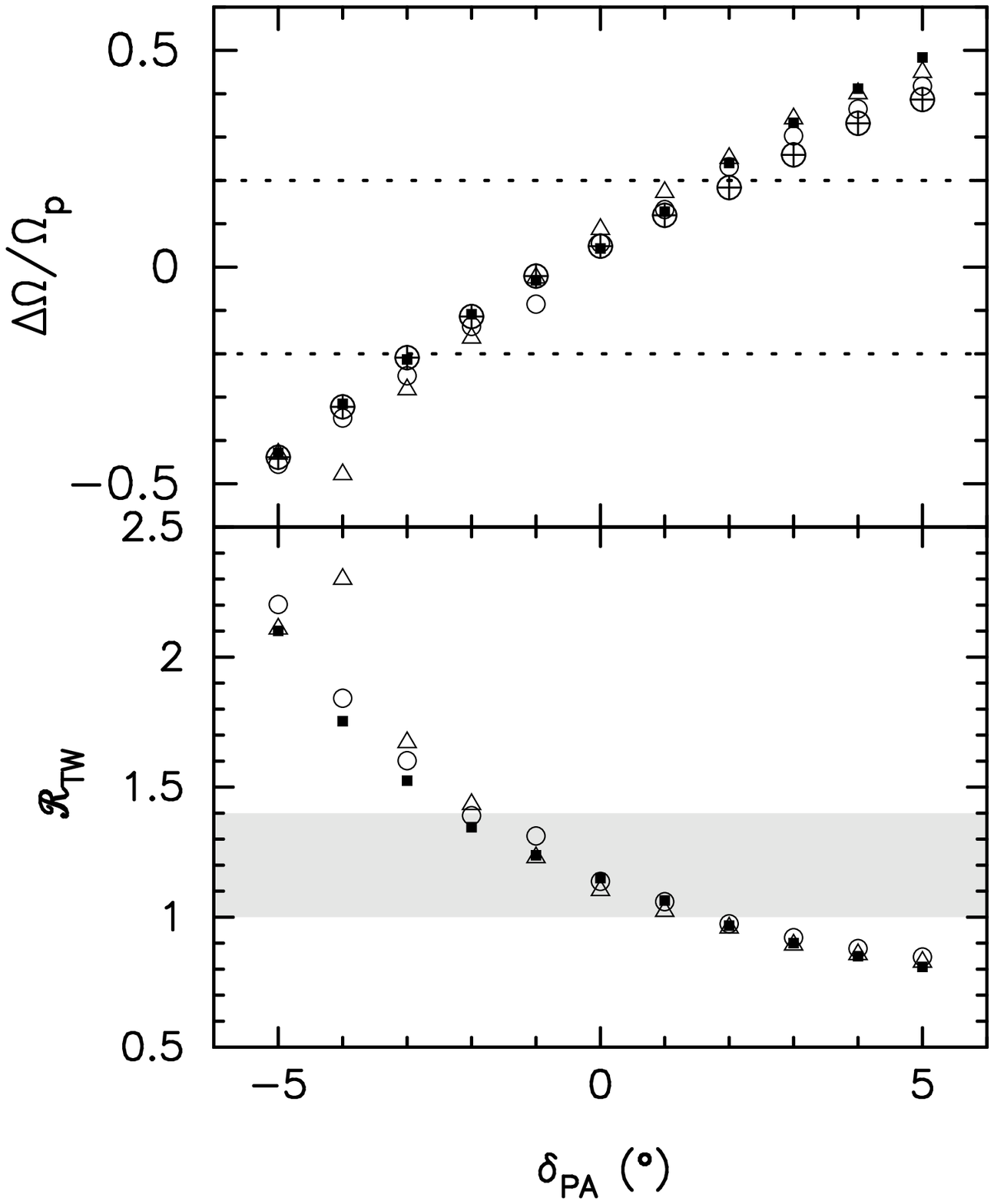,width=8.0truecm,angle=0}}
\caption[]{The variation of $\omerr$ (top) and $\vpdtw$ 
  (bottom) for small errors in \padisc, at $i=45\degrees$.  Circles, triangles
  and filled squares are for \pabar\ $= 30\degrees$, $45\degrees$ and
  $60\degrees$ respectively.  The dotted lines in the top panel represent
  errors of 20 per cent, while in the bottom panel, the shaded region
  indicates $1.0 \leq \vpdtw \leq 1.4$.  The crossed open circles in the top
  panel represent a system with $i$ and \pabar\ as in NGC 7079.}
\label{fig:vandperrs}
\end{figure}

\begin{figure}
\leavevmode{\psfig{figure=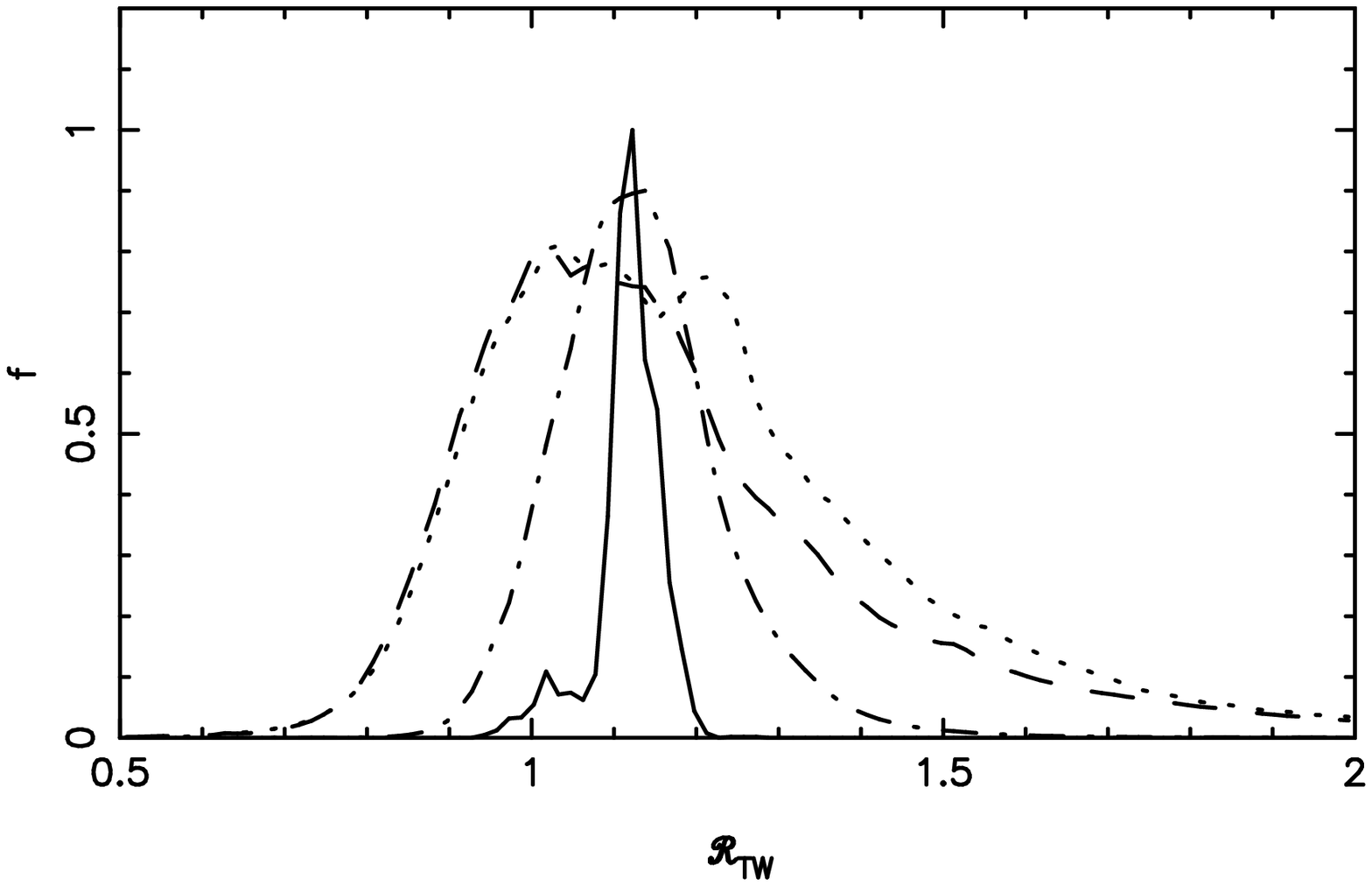,width=8.0truecm,angle=0}}
\caption[]{The distribution of $\vpdtw$ for various 
  distributions of random errors in \padisc.  The solid line shows the
  distribution without \padisc\ errors, while the dot-dashed and dashed lines
  show the distributions resulting from Gaussian errors of zero mean and FWHM
  = $2\degrees$ and $5\degrees$, respectively.  The dotted line is also for
  Gaussian errors with FWHM = $5\degrees$, but uses $\wtslit = \sigkin^{-2}$
  to measure $\omtw$.  Each line has been rescaled vertically for clarity.
  The distributions represent averages over $30\degrees \leq i \leq
  70\degrees$ and $10\degrees \leq$ \pabar\ $\leq 80\degrees$, and are not
  substantially changed by modest changes to these limits.}
\label{fig:distR}
\end{figure}

\subsection{Scatter from random \padisc\ errors}

Fig. \ref{fig:vandperrs} plots $\omerr$ and $\vpdtw \equiv (\om/\omtw) \vpd$
as functions of \paerr.  (This definition of $\vpdtw$ ignores the errors in
$V_c$ and $\len$ due to \paerr.  These errors change $\vpdtw$ by only a small
amount for the inclinations of interest here.)  The shaded region in the
bottom panel indicates the region of fast bars; it is clear that once
$|$\paerr$|$ becomes larger than about $2\degrees$, values of $\vpdtw$ scatter
outside this region.  Uncertainties in \padisc\ must therefore also contribute
to the scatter in measurements of $\vpd$.  Assuming Gaussian errors in
\padisc\ with zero mean and FWHM of $5\degrees$ ($2\degrees$), I found a
scatter in $\vpdtw$, $\scatterd$, (defined as the 67 per cent interval about
the median), of $\scatterd \simeq 0.4$ ($\scatterd \simeq 0.2$), as shown in
Fig. \ref{fig:distR}; this is substantially larger than the intrinsic
measurement scatter at \paerr\ $=0$, which is only $\scatter \simeq 0.06$.
Since, for the ADC sample, the observational root-mean-square uncertainty in
\padisc\ is $2\fdg1$, measurements of $\vpd$ with the TW method cannot
directly resolve the intrinsic distribution of $\vpd$ if it is as narrow as
hydrodynamical simulations require, even before other sources of scatter are
considered.

An important characteristic of the scatter is that $\vpdtw < 1$ may result.
Since $\vpd < 1$ is physically impossible (Contopoulos 1980), this may help in
distinguishing the effects of \padisc\ errors from the intrinsic distribution
of $\vpd$.

\section{Additional Non-Axisymmetries}
\label{sec:other_non_axi}

\begin{figure*}
\leavevmode{\psfig{figure=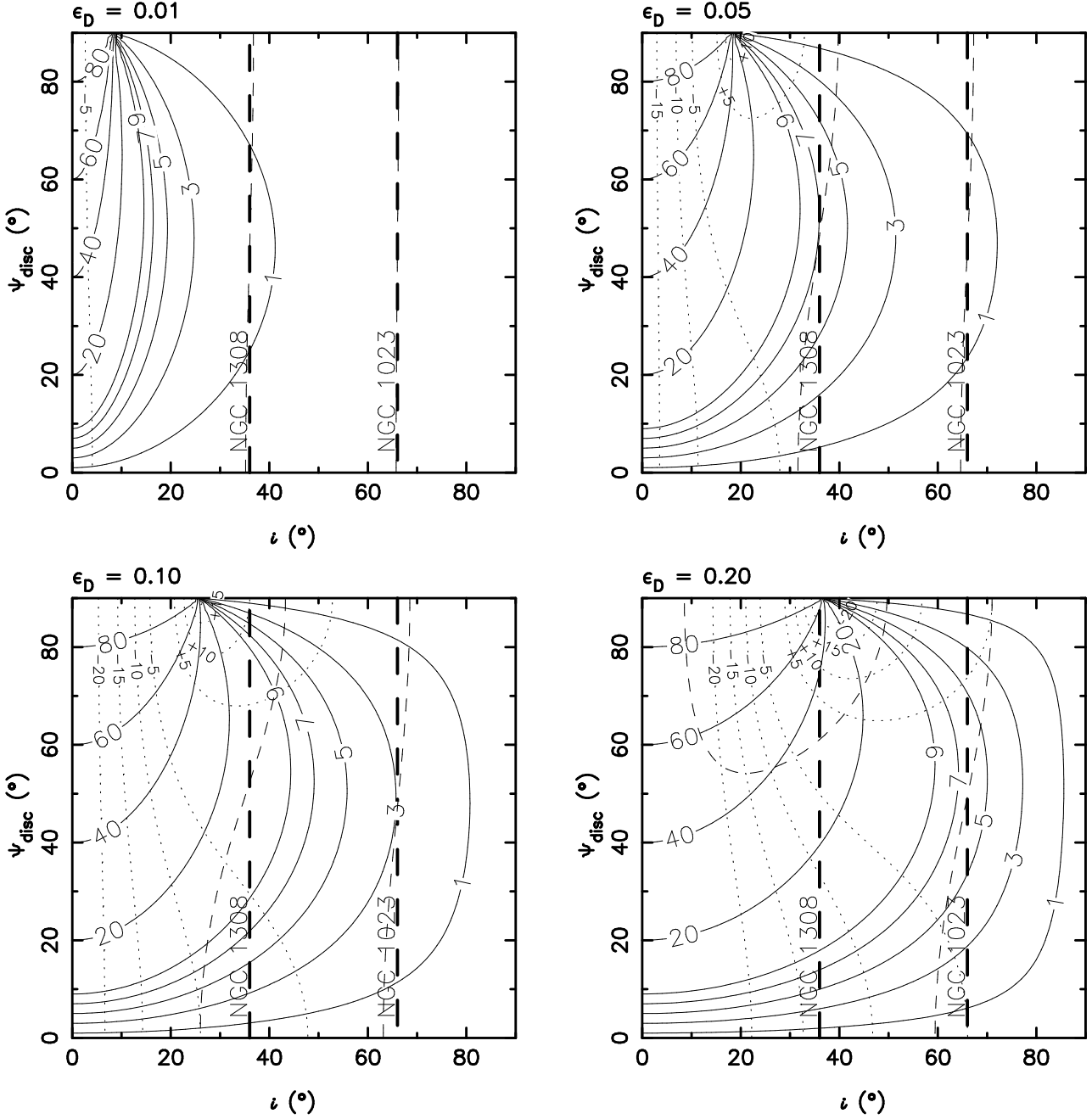,width=17.0truecm,angle=0}}
\caption[]{Contours of the errors in \padisc\ and $i$ resulting from assuming 
  that an intrinsically elliptical disc is circular.  The disc ellipticity,
  \epint, in each case is indicated in the top-left corner of each panel.  The
  solid contours show the errors in \padisc, while the dotted contours show
  the errors in $i$.  Each contour is labelled by the error it corresponds to;
  these are positive only for \padisc, because this figure only considers
  \paell\ $>0$, for the sake of simplicity.  For large inclinations (near
  edge-on), only very small errors in \padisc\ result, but as the disc becomes
  closer to face on, the errors generally become larger.  The dashed lines
  indicate the 2 galaxies on which the TW method has been used with the
  smallest (NGC 1308) and largest (NGC 1023) apparent inclination: the bold
  dashed lines are the inclinations assumed by the corresponding authors (see
  Table \ref{tab:tw_measurements}), which were obtained by assuming the outer
  disk is circular, while the thin dashed lines indicate the loci of
  $\epsilon_{app} = 1 - \cos i_{app}$.  Where \epint\ $> 1-\cos i$ (\eg\ NGC
  1308 when \epint\ $=0.2$), the typical errors in $i$ and, especially, in
  \padisc\ become very large, up to $90\degrees$.}
\label{fig:dpa}
\end{figure*}

If the disc contains additional non-axisymmetric structure besides the bar,
then this will interfere with the measurement of $\om$.  If the disc
non-axisymmetric density can be decomposed into 2 components, with different
pattern speeds, then $\omtw$ is a luminosity and asymmetry weighted average of
the two pattern speeds (Debattista \etal\ 2002b).  I assume that the second
component is a weaker non-axisymmetric structure and/or is at larger radius
and therefore lower surface brightness, so that this type of interference will
be relatively small and can be ignored.  (This can also be justified by noting
that the weak spiral structure at large radius in the $N$-body model does not
introduce substantial errors in $\omtw$.)  Instead, I concentrate only on the
effect these secondary non-axisymmetric structures have on $\omtw$ due to the
errors they introduce in the measurement of \padisc.

\subsection {Elliptical discs}
\label{sec:ell_discs}

In all cases in which the TW method has been used, \padisc\ has been measured
from surface photometry under the assumption that the disc is intrinsically
circular.  When the disc is elliptical, deprojecting with this assumption
gives rise to errors in $i$ and \padisc, as shown in Fig. \ref{fig:dpa}.
These errors lead to further scatter in $\vpdtw$.

\begin{figure}
\leavevmode{\psfig{figure=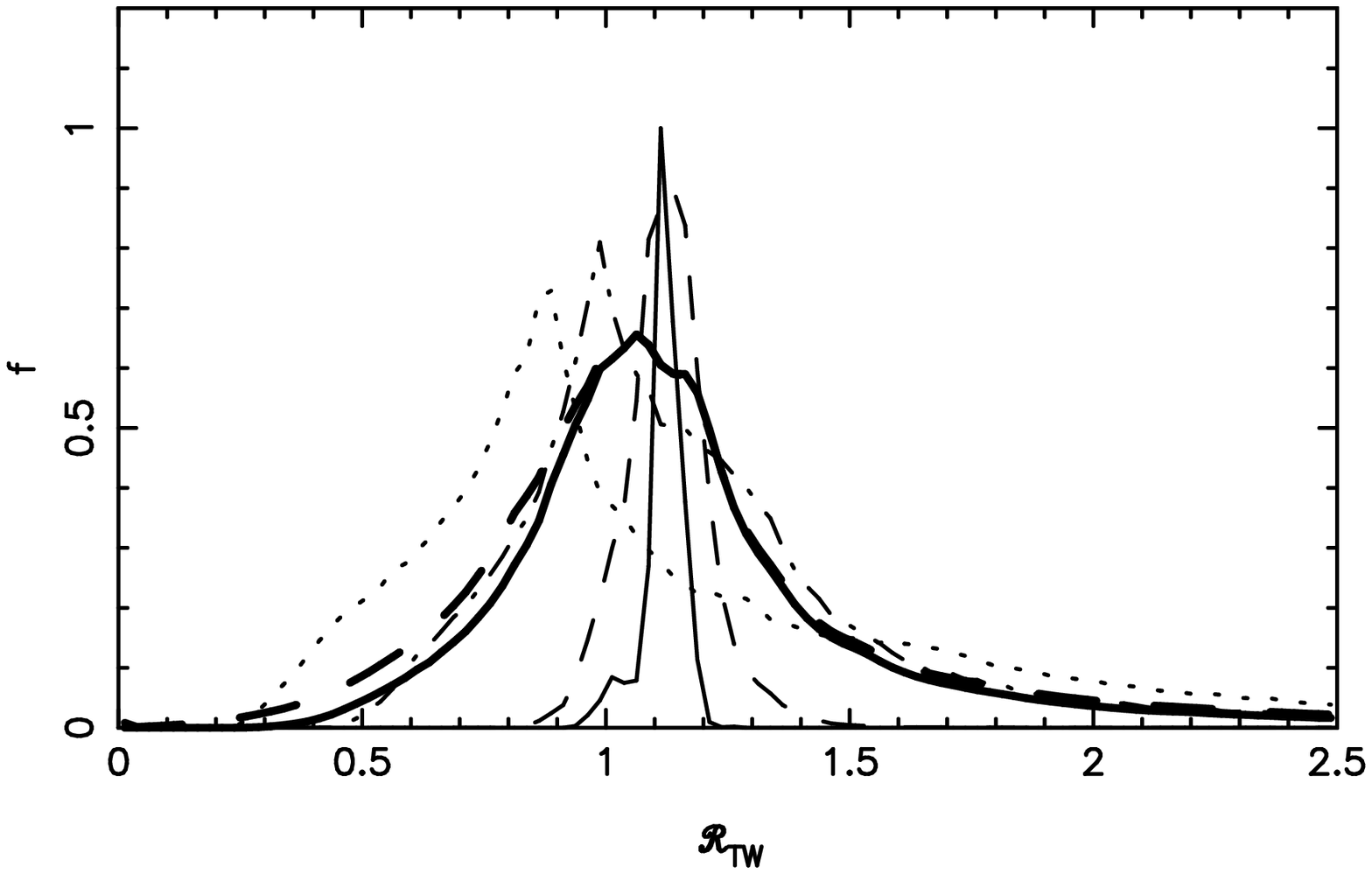,width=8.0truecm,angle=0}}
\caption[]{The distribution of $\vpdtw$ resulting from elliptical discs.  
  The thin solid line shows the intrinsic distribution when \epint\ $=0$,
  while the thin dashed, dot-dashed and dotted lines show the distributions
  resulting from errors caused by \epint\ $=0.01$, $0.05$ and $0.1$,
  respectively.  The thick solid and dashed lines are based on the empirical
  distribution of Andersen \& Bershady, with a maximum \epint\ of 0.1 and 0.15
  respectively.  Each line has been rescaled vertically for clarity.  The
  distributions represent averages over $30\degrees \leq i_{app} \leq
  70\degrees$ and $10\degrees \leq \psi_{b,app} \leq 80\degrees$, and are not
  substantially changed by modest changes to these limits.}
\label{fig:ellscatt}
\end{figure}

To study this scatter, I assumed that, at large radii, \epint\ and \paell\ 
(where \paell\ is the angle of the elliptical disc in the plane of the disc
relative to the line-of-nodes) are both constant, and computed the apparent
\padisc\ (PA$_{app}$) and apparent $i$ ($i_{app}$) resulting from the
assumption of a circular disc.  I used these to measure the apparent circular
velocity ($V_{c,app}$) and bar semi-major axis ($a_{B,app}$).  I then obtained
$\omtw \sin i_{app}$ as the slope of the best-fitting line to $(\pin,\kin)$,
from which I measured $\vpdtw = V_{c,app}/(a_{B,app}~ \omtw)$.  By assuming
that the bar is infinitely narrow, I measured the apparent bar PA in the disc
plane, $\psi_{b,app}$, and then averaged $\vpdtw$ over $30\degrees \leq
i_{app} \leq 70\degrees$, $10\degrees \leq \psi_{b,app} \leq 80\degrees$ and
$-90\degrees <$ \paell\ $\leq 90\degrees$.  Fig. \ref{fig:ellscatt} plots the
resulting distributions of $\vpdtw$ obtained for various constant \epint.  The
ellipticity-induced scatter, $\scattere$, grows rapidly with \epint\ 
($\scattere \simeq 0.2$, $0.6$ and $0.9$ for \epint\ $=0.01$, $0.05$ and $0.1$
respectively), with most measurements of $\vpdtw$ outside the range $1.0 \leq
\vpdtw \leq 1.4$ once \epint\ $=0.1$.  The distinctive peak to $\vpdtw < 1$
for the larger values of \epint\ is due to the fact that the distribution of
\paerr, at fixed $i_{app}$, has peaks near $\max(|$\paerr$|)$.  The peak at
$\vpdtw < 1$ is higher than that at $\vpdtw > 1$ because $\vpd \propto
\om^{-1}$.

Fig. \ref{fig:ellscatt} also shows the distribution of $\vpdtw$ resulting from
the ellipticity distribution of Andersen \& Bershady (2002) for later-type
unbarred galaxies.  The two largest values of \epint\ in their sample of 28
were \epint\ $=0.232^{+0.070}_{-0.064}$ and \epint\ $=0.165 \pm 0.083$
(Andersen 2002, private communication).  As can be seen in Fig. \ref{fig:dpa},
large values of \epint\ produce \padisc\ errors as large as $90\degrees$ in
the (apparent) inclination range of interest, which would result in very large
errors in $\omtw$.  Therefore I truncated their distribution at \epint\ $=0.1$
and $0.15$.  The resulting scatter is $\scattere \simeq 0.5$ and $\simeq 0.6$
respectively.

\begin{figure}
\leavevmode{\psfig{figure=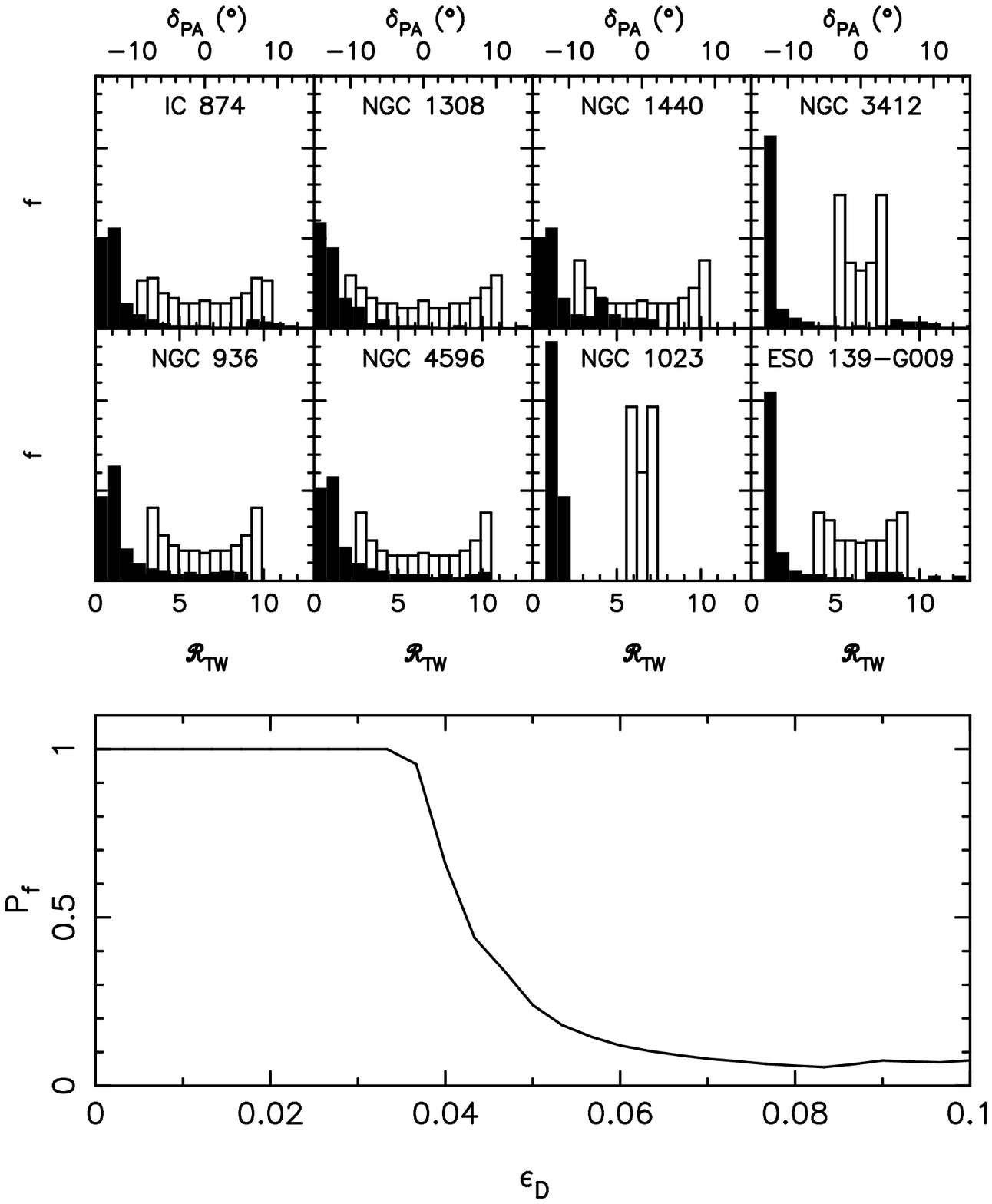,width=8.0truecm,angle=0}}
\caption[]{The bottom panel plots the probability, $P_f$, that none
  of the 8 galaxies of Table \ref{tab:tw_measurements} is outside the range
  $0.5 \leq \vpdtw \leq 2.5$ for \epint\ fixed for all galaxies.  The top 8
  panels show the distributions of $\vpdtw$ (filled histograms and bottom
  scale) and \paerr\ (open histograms, top scale) produced by matching
  $\psi_{b,app}$ and $i_{app}$ for the galaxies in Table
  \ref{tab:tw_measurements} under the assumption that \epint\ $=0.07$ (where
  $P_f < 0.1$).}
\label{fig:fast_probs}
\end{figure}

To compute an upper limit for the characteristic \epint\ of SB0 galaxies, I
define $P_f$ as the probability that all measurements will result in $0.5 <
\vpdtw < 2.5$, a range outside which, at the 67 per cent interval, none of the
measurements of Table \ref{tab:tw_measurements} fall.  Then, for that sample,
I compute $P_f$ by matching $i_{app}$ and $\psi_{b,app}$ to the observed
values and averaging over \paell, obtaining Fig. \ref{fig:fast_probs}.  The
probability of having found $\vpdtw$ less than $0.5$ or greater than $2.5$ for
one or more of these galaxies exceeds 90 per cent (75 per cent for $\vpdtw >
5.0$) if \epint\ $\geq 0.07$ for all of them.  (The strongest constraints come
from the low inclination galaxies, while NGC~ 1023, which has the largest
inclination of this sample, does not constrain \epint\ at all, up to $0.1$.)
This upper limit on the disc ellipticity is in rough agreement with previous
measurements (\eg\ Franx \& de Zeeuw 1992) for unbarred galaxies.

\subsection{Rings}

In Section \ref{sec:ell_discs}, I assumed that \paell\ is uncorrelated with
\pabar.  Correlations between \paell\ and \pabar\ may be introduced by the
outer rings often seen in SB galaxies.  Two main types of outer rings are
possible (\eg\ Buta 1995): $R_1$, which are aligned perpendicular to the bar,
and $R_2$, which line up with the bar.  Galaxies selected for TW measurement
do not contain strong rings, but conceivably weak rings might have been
overlooked.  To consider their effect on TW measurements, I simply set \pabar\ 
$=$ \paell\ (for rings of type $R_2$) and \pabar\ $=$ \paell $+ 90\degrees$
(for rings of type $R_1$) and proceeded as for Fig. \ref{fig:ellscatt}.  The
results, unsurprisingly, showed that rings of type $R_2$, which lead to
\paerr\ $\leq 0$, produce $\vpdtw \leq \vpd$, while rings of type $R_1$ lead
to $\vpdtw\geq \vpd$.  Buta (1995) found mean \epint\ of $0.26$ and $0.13 $
for rings of type $R_1$ and $R_2$ respectively.  If such rings had been
present in the sample of Table \ref{tab:tw_measurements}, then the scatter in
$\vpdtw$ would have been significantly higher.

\subsection{Spirals}

Recently, Barnes \& Sellwood (2003) have questioned the interpretation of
discrepancies between photometric and kinematic inclinations and PA's as
resulting from disc ellipticities.  Instead, they found evidence that spirals,
or similar non-axisymmetries, produce these discrepancies.  They reported an
average \padisc\ uncertainty of about $4\degrees$ for earlier-type galaxies.

\padisc\ errors of this type will produce scatter in $\vpdtw$ in much the same
way as do random \padisc\ errors.  For $\sigma \simeq 4\degrees$, I found a
resulting scatter $\scatters \simeq 0.7$.  However, the sample of galaxies
used by Barnes \& Sellwood (taken from Palunas \& Williams [2000]), excluded
galaxies as early as S0, so this value is somewhat uncertain and is probably
an over-estimate.

\subsection{Warps}

While most disc galaxies are coplanar inside $R_{25}$ (Briggs 1990), examples
of warps inside this radius are not unknown.  One extreme case is the
interacting galaxy NGC 3718, which has a warp of about $80\degrees$ at
$R_{25}$ (Schwarz 1985).  However, such strongly interacting galaxies are
usually not selected for TW studies.  Furthermore, the large velocity
dispersions of early-type galaxies serve to stiffen their stellar discs
(Debattista \& Sellwood 1999), so that any warps inside $R_{25}$ are generally
small.  Therefore warps probably do not introduce significant scatter in TW
measurements.

\section{Discussion and Conclusions}
\label{sec:discussion}

\subsection{How realistic are the error estimates?}

How realistic are these estimates of $\omerr$ and the $\scatter$'s?  Since,
for \paerr\ $<0$, the values of $\pin$ and $\kin$ are not all close to a
straight line (see Fig. \ref{fig:signals2}), a poor choice of $\wtslit$ could
lead to excess scatter.  At \paerr\ $=0$, I obtained the smallest $|\omerr|$
with $\wtslit = \sigkin^{-2}$, which is defined only from the variations of
$\kin$ with $X_{\rm max}$.  This is unsurprising, since $\sigkin$ represents
the full uncertainty in $\kin$.  All other definitions of $\wtslit$ produced
larger errors.  In particular, while $\wtslit = \sigkin^{-2}$ gives a mean
$\omerr$ of 3 per cent, $\wtslit = \stdwgts$ produces a mean $\omerr$ of 7 per
cent.

However, when \paerr\ $\neq 0$, $\wtslit$ favoring slits with small offset,
which generally acquire fractionally smaller perturbations, produces smaller
scatter.  Fig. \ref{fig:distR} compares the distributions of $\vpdtw$ from
random Gaussian \padisc\ errors of FWHM $=5\degrees$ as obtained using
$\wtslit = \stdwgts$ and $\wtslit = \sigkin^{-2}$.  The former produces a
smaller scatter, due mostly to the reduced noise at $\vpdtw > \vpd$, \ie\ at
\paerr\ $< 0$.  I tried other definitions of $\wtslit$, including
$\sigpin^{-2}$, equal weights, ${\cal P}$, and various combinations of these.
I also tried using only 3 slits (the central one and either the two with the
largest $|\pin|$ or the two flanking slits), as is often done in observations.
These always gave larger scatter, typically by $20$ per cent or more.  I
therefore used $\wtslit = \stdwgts$ everywhere in this paper to compute
$\omerr$ and the $\scatter$'s.  Thus I am assured of a conservative estimate
of the scatter, while also matching better the main source of noise in the
observations: the photon statistics.

Since I have used only one simulation to estimate the scatter, I need to show
that this simulation does not over-estimate the errors in $\omtw$ that real
galaxies would suffer.  Perhaps the most important parameter affecting the
size of the scatter in $\omtw$ is $\len/R_{\rm d}$, as described in Section
\ref{sec:tw_method}.  A series of experiments with razor-thin, flat rotation
curve, axisymmetric exponential discs showed that, indeed, the scatter in
$\omtw$ due to random \padisc\ errors increases as $R_{\rm d}$ decreases.
Since my model SB0 has a value of $\len/R_{\rm d}$ that is towards the upper
end of those in the ADC sample, my measurements of $\scatterd$ and $\scattere$
probably underestimate somewhat the scatter which the same \padisc\ errors
would produce in real galaxies.  The same conclusion resulted from a test with
a lower quality ($102K$ particle) simulation having a larger bar ($\len/R_{\rm
  d} = 2.6$); for random Gaussian errors of FWHM $=5\degrees$, this bar
produced $\scatterd = 0.3$ versus $0.4$ for the shorter bar used in this
paper.

The trend with \paerr\ seen in Fig. \ref{fig:signals2} is in the same sense as
was found by Debattista \& Williams (2002) for NGC 7079.  Fig.
\ref{fig:vandperrs} plots $\omerr$ for the same projection as NGC 7079.  The
errors in $\omtw$ due to \paerr\ for NGC 7079 ($\len/R_{\rm d} = 1.5 \pm 0.2$)
reported by Debattista \& Williams are perhaps a little larger than those
computed here.  Gratifyingly, the error estimates produced by the $N$-body
model are not unrealisticly large.

\subsection{The ellipticity of early-type barred galaxies}

The ellipticities of S0 galaxies are poorly constrained.  From photometry
only, Fasano \etal\ (1993) found that they could not rule out that they are
perfectly oblate.  The two S0 galaxies with directly measured ellipticities,
IC 2006 (Franx \etal\ 1994) and NGC 7742 (Rix \& Zaritsky 1995) both have
small, possibly zero, ellipticity ($\epsilon_\Phi = 0.012 \pm 0.026$ and $0.02
\pm 0.01$ respectively).  The ellipticities of SB galaxies are not much better
constrained, undoubtedly because they require a distinction between the inner,
bar-dominated, region and the outer parts.  Photometry alone, therefore, is of
limited use, and kinematics also are needed.  Unfortunately, most TF studies
have avoided SB galaxies.  Debattista \& Sellwood (2000) showed that the small
fraction of bright ($M_I \leq -21$) SB galaxies contaminating the sample of
Mathewson \& Ford (1996), who selected against SB galaxies, satisfies the same
TF relation, and has the same scatter, as the unbarred (SA) galaxies.  Sakai
\etal\ (2000) calibrated the TF relation of nearby galaxies with Cepheid
distances; their sample of 21 galaxies contained a more representative
fraction of SB galaxies, at $\sim 30$ per cent.  The resulting TF relation,
including the scatter, also was identical for SA and SB galaxies.  Thus we may
suppose that the TF-based constraint of Franx \& de Zeeuw (1992),
$\epsilon_\Phi < 0.1$, also holds for SB galaxies.

The constraint obtained here, \epint\ $\ltsim 0.07$, is in rough agreement
with the constraints for SA galaxies.  However, an important possible bias
needs to be pointed out.  The ADC sample of 6 galaxies explicitly excluded
galaxies for which, at large radius, the observed \padisc\ changes
substantially with radius.  From a sample of 11 galaxies for which they
obtained surface photometry, one (Aguerri 2002, private communication) was
excluded for this reason.  If either \epint\ or \paell\ changes with radius,
then the observed changes in \padisc\ will typically be greater in galaxies
with larger mean \epint.  Thus the cut on the size of \padisc\ variations may
have introduced a bias in the ellipticity distribution of the ADC sample; on
the other hand, large variations in \padisc\ may have been caused instead by
spirals or by a warp.

Although these constraints on SB galaxy ellipticities are consistent with the
constraints on SA galaxy ellipticities, this does not mean that their
ellipticity distributions are the same, since both the TF and the TW
constraint obtain only upper limits on \epint.

\subsection{The intrinsic distribution of $\vpd$}

Hydrodynamical simulations of SB galaxies find a narrow range of $\vpd = 1.2
\pm 0.2$.  The presently observed distribution of $\vpdtw$ is dominated by the
observational uncertainties in $\omtw$, $\len$ and $V_c$.  Nevertheless, it is
clear that all 8 galaxies measured so far are consistent with the range found
in hydrodynamical simulations.  In their $N$-body simulations with
cosmologically motivated initial conditions, Valenzuela \& Klypin (2002) found
that bars with $\vpd = 1.7$ were produced, which they considered to be
consistent with the observations.  Indeed, for 4 of the 8 galaxies listed in
Table \ref{tab:tw_measurements}, $\vpd = 1.7$ is within the error interval.
However, 3 of these 4 galaxies are the ones with the largest error bars, and
the fourth galaxy is only just barely consistent with this value.  For the ADC
sample, which have well-determined \padisc\ uncertainties, the
root-mean-square uncertainty in \padisc\ is $2\fdg1$.  From the results of
Section \ref{sec:tw_method}, the corresponding scatter in $\vpdtw$, excluding
any contribution due to disc ellipticity, should be $\scatterd \simeq 0.4$.
Allowing for this scatter, it seems possible that $1.7$ is outside the
intrinsic range of $\vpd$.

For a crude estimate of the intrinsic range of $\vpd$, suppose we can write
$\Delta_{\vpd, obs}^2 = \Delta_{\vpd, int}^2 + \Delta_{\vpd, \delta}^2 +
\Delta_{\vpd, \epsilon}^2 + \Delta_{\vpd, unc}^2$, where $\Delta_{\vpd, obs}$
is the observed scatter, $\Delta_{\vpd, int}$ is the intrinsic range of
$\vpd$, $\Delta_{\vpd, \delta}$ is the scatter due to random \padisc\ errors,
$\Delta_{\vpd, \epsilon}$ is the scatter due to disc ellipticity and
$\Delta_{\vpd, unc}$ is the scatter induced by uncertainties in the
measurements of $\om$, $\len$ and $V_c$.  All these $\scatter$'s are assumed
to be 67 per cent intervals.  (Other sources of scatter, such as direct
interference from spiral or other structure, small errors in slit orientation,
\etc, may be present but are assumed here to be unimportant.)  From Section
\ref{sec:tw_method} I get that $\Delta_{\vpd, obs} \simeq 1.0$ and
$\Delta_{\vpd, unc} \simeq 0.7$, while from Section \ref{sec:randomerrs} I get
$\Delta_{\vpd, \delta} \simeq 0.4$.  If \epint\ $=0$ for all galaxies, then
$\Delta_{\vpd, int} \simeq 0.6$, while the distribution of \epint\ of Andersen
\& Bershady (2002), truncated at \epint\ $=0.1$, produces $\Delta_{\vpd, int}
\simeq 0.3$.  If, on the other hand, the interpretation of Barnes \& Sellwood
(2003) is correct, then $\scattere = 0$, but it is replaced by $\scatters
\ltsim 0.7$.  It therefore seems possible that the intrinsic range of $\vpd$
for early-type galaxies spans a range similar to the later-type galaxies.

Unfortunately, the sample size is still too small for a proper statistical
test of this suggestion.  If correct, then the fact that SB galaxies have the
same distribution of $\vpd$ as the more gas-rich later-type SB galaxies
requires that gas is not dynamically very important for the evolution of
$\om$.

\subsection{Future work and conclusions}

The current sample of TW measurements is still quite small, so it is not
unlikely that, in the future, more measurements will be obtained.  The results
of this paper can be read as an endorsement of careful surface photometry of
target galaxies to accurately measure \padisc.  Inclinations in the range
$50\degrees \leq i \leq 60\degrees$ are preferable, since they are less
sensitive to errors in \padisc.  For statistical studies, especially to
constrain the distribution of $\vpd$, it would be very useful if future
studies were to report their uncertainty in \padisc.  Galaxies with strong
outer rings do not make good candidates for TW measurement because of the
inherent uncertainty in \padisc, and should be avoided.  If the TW method is
ever to be used on late-type galaxies, perhaps in the infra-red (\eg\ Baker
\etal\ 2001), care must be taken that the presence of spirals does not lead to
excess errors in \padisc.

The pattern speed of triaxial elliptical galaxies is a matter of theoretical
speculation.  Because of the large velocity dispersions and low stellar
streaming velocities, it is generally thought that their pattern speeds must
be small.  Measurement of their pattern speeds would be very interesting, but
unfortunately, application of the TW method to elliptical galaxies is likely
to be accompanied by significant uncertainty in their intrinsic orientations
(amongst other difficulties).  Thus TW measurements of their pattern speeds
may have large uncertainties.

I have shown that errors in \padisc\ lead to significant error in TW
measurements.  For the observational level of random Gaussian errors, the
resulting scatter in $\vpd$ is $\scatterd \simeq 0.4$.  If barred galaxies are
also modestly elliptical, then the total scatter increases further, depending
on the distribution of \epint.  Given the observed range of $\vpd$, this
suggests, therefore, that the gas-poor early-type galaxies have a narrow
distribution of $\vpd \sim 1.0 - 1.4$, not much different from gas-rich
late-type galaxies, as determined by independent means.  This result would
imply that gas is not dynamically important for the evolution of bar pattern
speeds.

\bigskip
\noindent
{\bf Acknowledgments.} 

\noindent
This work has been made possible by support from the Schweizerischer
Nationalfonds through grant 20-64856.01.  I thank Enrico Maria Corsini, Joris
Gerssen, Dave Andersen, Jerry Sellwood, Alfonso Aguerri and Niranjan Sambhus
for comments and discussion.

\end{document}